\documentclass[journal=jpclcd,manuscript=article]{achemso}
\usepackage{chemformula} 
\usepackage[T1]{fontenc} 
\usepackage{mathtools}
\usepackage{amsthm}
\usepackage{amssymb}
\usepackage{bm}
\usepackage[version=3]{mhchem}
\usepackage{braket}
\usepackage{relsize}
\usepackage{enumitem}
\usepackage[vlined]{algorithm2e}
\usepackage{algpseudocode}
\usepackage{graphicx}
\usepackage{footnote}
\usepackage{multirow}
\usepackage{tabularx}
\usepackage{booktabs}
\usepackage{makecell}
\usepackage{threeparttable}
\usepackage{color}
\usepackage{breakurl}
\usepackage{epstopdf}

\newcolumntype{C}{>{\centering\arraybackslash}X}


\title{The Doubles Connected Moments Expansion: A Tractable Approximate Horn-Weinstein Approach for Quantum Chemistry}

\author{Brad Ganoe}
\author{Martin Head-Gordon}
\affiliation{
Pitzer Center for Theoretical Chemistry, Department of Chemistry,  University of California, Berkeley CA 94720
}
\email{ganoe@berkeley.edu}
\begin{document}

\begin{tocentry}
\centering
\includegraphics[width=\linewidth]{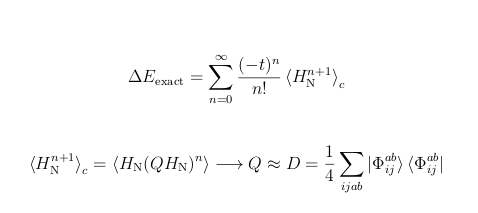}
\label{toc_graph}
\end{tocentry}

\begin{abstract}
\textit{Ab initio} methods based on the second-order and higher connected moments, or cumulants, of a reference function have seen limited use in the determination of correlation energies of chemical systems throughout the years. Moment-based methods have remained unattractive relative to more ubiquitous methods, such as perturbation theory and coupled cluster theory, due in part to the intractable cost of assembling moments of high-order and poor performance of low-order expansions. Many of the traditional quantum chemical methodologies can be recast as a selective summation of perturbative contributions to their energy; using this familiar structure as a guide in selecting terms, we develop a scheme to approximate connected moments limited to double excitations. The tractable Double Connected Moments (DCM(N)) approximation is developed and tested against a multitude of common single-reference methods to determine its efficacy in the determination of the correlation energy of model systems and small molecules. The DCM(N) sequence of energies exhibits smooth convergence towards limiting values in the range of $\mathrm{N}=11-14$, with compute costs that scale as a non-iterative $O(M^{6})$ with molecule size, $M$. Numerical tests on correlation energy recovery for 55 small molecules comprising the G1 test set in the cc-pVDZ basis show that DCM(N) strongly outperforms MP2 and even CCD with a Hartree-Fock reference. When using an approximate Brueckner reference from orbital-optimized (oo) MP2, the resulting oo:DCM(N) energies converge to values more accurate than CCSD for 49 of 55 molecules. The qualitative success of the method in regions where strong correlation effects begin to dominate, even while maintaining spin purity, suggests this may be a good starting point in the development of methodologies for the description of strongly correlated or spin-contaminated systems while maintaining a tractable single-reference formalism.
\end{abstract}

\section{Introduction}
The majority of quantum chemical methods begin from the Schr\"{o}dinger equation and make approximations in an effort to efficiently find an acceptably accurate expectation value of the Hamiltonian $E = \braket{\phi|H|\phi}$. As the approximate wavefunction $\ket{\phi}$ approaches the exact ground state wavefunction $\ket{\psi_0}$, our approximate expectation value approaches the true energy. Fine tuning this approach takes many forms, but generally begins from a reference wavefunction such as the Hartree-Fock determinant or the antisymmetrized product of strongly orthogonal geminals, and then determining the correlation energy of the system to approach the exact energy or values that are within 'chemical accuracy' for various systems. Popular among these are methods based around some variant of perturbation theory, in which the correlation energy of the system:
\begin{equation}
\Delta E_{\text{exact}}=\braket{0|V_{\text{N}}|\psi_0} = \sum\limits_{m=0}^{\infty}\braket{0|V_{\text{N}}[R_{0}V_{\text{N}}]^{m}|0}
\end{equation} 
is expanded in a series approximation, with $V_{\text{N}}$ being the normal-ordered perturbation operator and $R_{0}$ being a resolvent operator based on the expansion. Different partitionings and reference functions give rise to well-known perturbation series expansions\cite{mp_1}, such as Brillouin-Wigner\cite{mp_2}, Rayleigh-Schr\"{o}dinger\cite{mp_4}, or M\o ller-Plesset\cite{Moller:1934} if performed with a Hartree-Fock reference. In the latter case, the resolvent is then simply $R_0 = \left(F-\braket{0|F|0}\right)^{-1}$, where $F$ is the mean field Fock operator.

The Coupled Cluster expansion\cite{cc_1,cc_2,cc_3,cc_4} may likewise be obtained through the use of an exponential wavefunction ansatz and its projection onto a reference and some set of substituted or excited determinants. 
One may also explore the Hilbert space variationally such as in the CI, CASSCF\cite{cas} and DMRG\cite{dmrg} schemes. What all these methods have in common is that the expectation value of the Hamiltonian is the primary quantity of interest. It is interesting to note that additional expectation values implicitly arise in many of these, such as the variance, or second moment, of the energy:
\begin{equation}
\mu_2 = \braket{H^{2}}-\braket{H}^{2}
\end{equation}
In the variational methods, the variance approaches zero as the expectation value of $\braket{H}$ approaches the exact energy (i.e. as $\ket{\phi}$ approaches the exact eigenstate, $\ket{\psi}_0$). The variance also becomes an important quantity in the coupled cluster equations,\cite{Small:2007} which can be made apparent by inserting the identity in the variance expression 
\begin{equation}
\mu_{2} = \braket{H(P+Q)H} - \braket{HPH} = \braket{HQH} = \braket{H^{2}}_{\text{c}}  
\end{equation} 
We use $P$ to mean the projective subspace of the reference function and $Q$ its orthogonal subspace such that $P+Q=I$ and $\mu_{2} = \braket{H^{2}}_{\text{c}}$ to mean the second cumulant or equivalently the second connected moment of the Hamiltonian. The amplitude equations are solved using $Q\bar{H}P=0$, where $\bar{H}$ is the similarity-transformed Hamiltonian, which  results in a value of $\mu_{2}=0$ within the space, $Q$, within which the cluster equations are solved.

Moments play a central role in an alternate approach, the so-called t-expansion of Horn and Weinstein\cite{texp}. This approach begins from the observation that a given approximate ground state wavefunction, $\ket{\phi}$ can always be improved by defining:
\begin{equation}
    \ket{\psi (t)} = \frac{ e^{-tH/2} \ket{\phi}} { {\braket{\phi|e^{-tH}|\phi}}^{1/2} }
 \label{eq:improved}
\end{equation}

\begin{equation}
    E(t) = \braket{\psi(t)|H|\psi(t)} = \frac{ \braket{\phi|He^{-tH}|\phi} } {\braket{\phi|e^{-tH}|\phi}} = -\frac{Z'(t)}{Z(t)}
\end{equation}
Here $Z(t)$ is a Maclaurin series containing expectation values of powers of the Hamiltonian operator, $\braket{H^j} = \braket{\phi|H^j|\phi}$:
\begin{equation}
Z(t) = \braket{\phi|e^{-tH}|\phi} = \sum\limits_{j=0}^{\infty}\frac{(-t)^{j}}{j!}\braket{H^j}
\end{equation}
$Z'(t)$ is defined as:
\begin{equation}
    Z'(t) = \frac{dZ(t)}{dt} = - \sum\limits_{j=0}^{\infty}\frac{(-t)^{j}}{j!}\braket{H^{j+1}}
\end{equation}
Thus the expression for $E(t)$ above is the logarithmic derivative of $Z(t)$. The result for the energy can be generalized to other operators, $O$, to yield an  expression for $O(t)$ which is likewise exact in the limit as $t \rightarrow \infty$.\cite{texp}

The promise of the $t$-expansion is that for an arbitrary reference $\ket{\phi}$ satisfying $\braket{\phi|\Psi}\neq0$, the exact energy cannot only be approached by the conventional variational approach of treating $\braket{H}$ as accurately as possible by improving the trial function, $\ket{\phi}$. Instead, exactness can also be attained by deriving an energy expression in terms of coefficients, $\mu_j$ that involve powers of the Hamiltonian, evaluated with a fixed approximate wavefunction:
\begin{equation}
    E(t) = \sum\limits_{j=0}^{\infty}\frac{(-t)^{j}}{j!} \mu_{j+1}
    \label{eq:connected_moments}
\end{equation}
Simple powers of the Hamiltonian, $\braket{H^j}$, scale as powers of the system volume, $V^j$, whilst of course the energy is linear in the volume (extensive). Thus the coefficients, $\mu_j$ that enter the expansion above, defined by the negative quotient of the two power series, $Z'(t)$ and $Z(t)$, must also scale linearly in the volume.  These $\mu_j$ coefficients are the \textit{connected  moments}, $\mu_j = \braket{H^j}_c$, which may be defined in the diagrammatic sense of connectedness, or equivalently by evaluating the power series quotient to obtain the $c_j$ term by term. Thus $\mu_1 = \braket{H}$, $\mu_2 = \braket{H^2} - \braket{H}^2$, $\mu_3 = \braket{H^3} - \braket{H}\braket{H^2} - \braket{H^2}_c \braket{H}$, etc, and in general:
\begin{equation}
    \mu_j \equiv \braket{H^j}_c = \braket{H^j} - \sum\limits_{k=0}^{j-1} \binom{j}{k} \mu_{k+1} \braket{H^{j-k}}
\end{equation}


Turning the $t$-expansion into a practical computational method is a non-trivial challenge. Creating regular expressions for this series has resulted in various extrapolations\cite{texp2,pade}. At low order, amongst the most promising of these is the Connected Moment Expansion (CMX)\cite{cmx1,cmx2}. A useful form of the CMX can obtained by presuming that an $N^\text{th}$ order approximation to $E(t)$, Eq. \ref{eq:connected_moments}, can be expressed as a sum of $(N-1)$ decaying exponentials, $A_i e^{-b_it} $, and obtaining the coefficients by matching low order terms in the power series expansions.\cite{cmx2} In the resulting CMX models, an $N^{\text{th}}$ order  approximation, CMX($N$) to the ground state energy may be succinctly written\cite{cmx2} in terms of the lowest $2N-1$ connected moments, and the inverse of an $(N-1) \times (N-1)$ matrix constructed from those moments:
\begin{equation}
    E^{(N)} = \mu_{1} - 
\begin{bmatrix}
\mu_{2}&\mu_{3}&\cdots&\mu_{N}
\end{bmatrix}
\begin{bmatrix}
\mu_{3}&\mu_{4}&\cdots&\mu_{N+1}\\
\mu_{4}&\mu_{5}&\cdots&\mu_{N+2}\\
\vdots&\vdots&\ddots&\vdots\\
\mu_{N+1}&\mu_{N+2}&\cdots&\mu_{2N-1}\\
\end{bmatrix}^{-1}
\begin{bmatrix}
\mu_{2}\\
\mu_{3}\\
\vdots\\
\mu_{N}
\end{bmatrix}
\label{eq:CMX(N)}
\end{equation}
In particular, CMX(1) is simply the energy of the reference (i.e. $\mu_1)$, and the two lowest order corrections to the reference energy, CMX(2) and CMX(3), have energies defined as 
\begin{equation}
E_{\text{CMX(2)}}=\mu_{1} - \frac{\mu_{2}^{2}}{\mu_{3}}
\label{eq:CMX(2)}
\end{equation}
and
\begin{equation}
E_{\text{CMX(3)}}=\mu_{1} - \frac{\mu_{2}^{2}}{\mu_{3}}-\frac{\mu_{3}(\mu_{4}\mu_{2}-\mu_{3}^{2})}{\mu_{5}\mu_{3}-\mu_{4}^{2}}
\label{eq:CMX(3)}
\end{equation}
respectively. CMX(4) requires moments up to $\mu_7$ and the inverse of a $3 \times 3$ moment matrix. Several variants\cite{amx,gmx} have also been developed to help rid the original expansion of singularities, and combine the terms in different combinations. 

The $t$-expansion is not the only place these connected moments have been the central quantity in determining the energy. L\" owdin's implicit energy formula\cite{lowdin}
\begin{equation}
    1 = \braket{\Phi|H(E-PH)^{-1}|\Phi}
\end{equation}
may be expanded\cite{con_denom} using the identity
\begin{equation}
    (\hat{A} - \hat{B})^{-1} = \hat{A}^{-1} + \hat{A}^{-1}\hat{B}(\hat{A} + \hat{B})^{-1}
\end{equation}
to obtain an approximate expression for the roots of the energy:
\begin{equation}
    E^{(N)} = \mu_{1}E^{(N-1)} + \mu_{2}E^{(N-2)} + \cdots + \mu_{(N-1)}E^{(1)} + \mu_{N}
\end{equation}
This set of equations requires the $N$ lowest connected moments to evaluate the $N^{\text{th}}$-order approximation to the energy in the denominator-free perturbation theory.\cite{con_denom}. At second order the result is $E^{(2)}=\mu_1 + (\mu_2)^2/\mu_1$. If one optimizes the constant energy denominator associated with the first order wavefunction, the CMX(2) result is recovered at second order, with the third order correction going to zero.


The CMX-based moments approaches and alternatives have been investigated across both model \cite{mancini1991approximations,fessatidis2008zero,pade,mancini1994numerical,fessatidis2010analytic,cioslowski1987connected,cioslowski1987study,cioslowski1987estimation,cioslowski1987connected2,lo1994connected,fernandez2009rayleigh,fernandez2009perturbation}\cite{,cioslowski1987study,cioslowski1987estimation,cioslowski1987connected2,lo1994connected,fernandez2009rayleigh,fernandez2009perturbation} and real \cite{mancini1991approximations,mancini1994numerical,amore2012high,fessatidis2002moments,gmx,massano1989application,mancini1991ground,ZHURAVLEV20161995} systems in lattice gauge theory, quantum chromodynamics, and quantum chemistry, which has been well reviewed by Amore\cite{Amore_2011}. The convergence and analytic behaviors of these methods have been the focus of several studies in the literature and their strengths and limitations have been well evaluated\cite{lee1993application,mancini1994analytic,mancini1994numerical,ullah1995removal,amore2012high,amore2013solution}. Focusing on the problem of the correlation energy of molecular systems, moment-based methods have remained a niche topic of research relative to standard coupled cluster or many-body approaches based on the Schr\"{o}dinger equation, with only a small catalogue of papers using moments-based methods for electronic structure\cite{cioslowski1987connected,cmx2, con_denom,mancini2005generalized,yoshida1988connected,noga2002use}. Of particular note is the Method of Moments Coupled Cluster (MMCC) approach\cite{kowalski2000method, piecuch2002recent}, which acknowledges that the exact energy may be retrieved from any Coupled Cluster reference with energy $E_{CC}$ via the formalism of $\beta$-nested equations by projecting onto the asymmetric energy expression
\begin{equation}
    \delta E = E - E_{CC} = \frac{\braket{\Psi_{0}|(H-E_{CC})e^{T}|0}}{\braket{\Psi_{0}|e^{T}|0}}
    \label{eq:piotr}
\end{equation}
This forms an approximate hierarchy of non-iterative corrections yielded through better approximations to $\ket{\Psi_0}$ by including terms related to the excitation space neglected in the Coupled Cluster amplitudes, but are extant in the general moment contributions to the exact energy\cite{pimienta2003method}. For example, use of a CCSD wavefunction in MMCC requires only up-to hextuply-excited configurations; while the triple excitations and higher do not factor into the amplitude equations and/or second moment, they do survive in the exact representation of Equation \ref{eq:piotr}.

One reason for the dearth of investigation is the cost to assemble these connected moments and the order to which one needs them to approach quantitative accuracy. Early correlation energy studies\cite{cmx2,cioslowski1987molecules} compared the lowest order CMX expansions to the similar M\o ller-Plesset series. In weakly correlated systems, the CMX(2) underperforms the MP2 energy whilst requiring greater computational costs. Assembling $\mu_{3}$ requires $O(N^{6})$ tensor contractions, which is the most costly step in the evaluation of CMX(2) with a Hartree-Fock reference. In fact, CMX(2) is fully equivalent to the Uns\" old approximation to MP2\cite{con_denom}, which helps explain this underperformance in the weakly-correlated region. Despite this quantitative failure, some successes have been reported in the strongly-correlated regions of bond breaking, where many traditional single-reference quantum chemical methods break down or become prohibitive in cost and the Uns\"old approximation does not diverge, unlike the traditional MP2 approximation. Going to the next order, CMX(3) requires the assembly of $\mu_{5}$, requiring $O(N^{8})$ operations which is already cost-prohibitive for molecules of medium-size (the same scaling as full CCSDT, for example). We also note that there has been recent interest in the connected moments as an approach for performing molecular electronic structure calculations on quantum computers.\cite{kowalski2020quantum,claudino2021improving}

Efforts to truncate the many-body expansion of the Full CI (FCI) problem in a computationally tractable way has led to many different ansatze to approximate the ground-state energy of the system. The most successful quantum chemical approaches typically cover subset of the FCI expansion. An order-by-order truncation in the fluctuation parameter results in the regular MP(n) expressions, but more sophisticated summations are possible. RPA, for example, is equivalent to an infinite sum over the ring diagrams in the perturbation expansion\cite{Ren2012}. Infinite-order resummations of the perturbative terms also result in various coupled cluster methodologies. The MBPT series summed over all possible doubly excited states, DMBPT($\infty$)\cite{bartlett1977comparison}, is formally equivalent to CEPA(0)\cite{Cizek1971359} and linearized CCD\cite{vcivzek1966correlation}. This perturbative analysis\cite{raghavachari1990fifth} holds true for various CC expansions\cite{ccsd,ccsdt} as well, with a notable example being CCSD(T)\cite{pT_1,pT_2} as all connected singles and doubles to infinite-order, as well as triples through fourth-order with a perturbative fifth-order correction of the triple's contribution to the single's amplitudes\cite{STANTON1997130}. 
Several approximations reliant on removing terms in the series have arisen and has been well reviewed recently\cite{Varun}.

Despite the success of these selective summations in perturbation theory, no comparable scheme has been reported in approximating connected moments for use in the CMX-based methodologies reliant on their use. Keeping in mind what has been successful in the perturbative case, we report a methodology to approximate the connected moments using a selective summation constrained to the doubly-excited manifold and apply these approximate connected moments in place of their vastly more expensive exact counterparts. We design a recursive algorithm to construct approximated connected moments, and use these terms to assess the validity in their use for approximating correlation energies across a range of small molecules and report stable results up through twentieth order in the CMX expansion.

\section{The Doubles Connected Moment Approximation}

To begin evaluating finite-order CMX approximations we must specify a reference from which to construct the connected moments. In this work, we shall take both the Hartree-Fock determinant, $\ket{\phi} = \ket{\Phi_{HF}}$, and the SCF reference generated via orbital optimized (OO) MP2,\cite{Lochan:2007:O2,Lee:2018a,rettig2022revisiting} $\ket{\phi} = \ket{\Phi_{OOMP2}}$ as our zeroth-order wavefunctions. The latter is a simple approximation to the exact Brueckner determinant which is best single reference.\cite{handy1989size,raghavachari1990size} Use of only the occupied molecular orbitals in the reference serves us both as a simple model from which to understand the behavior of approximate moment methods in relation to traditional single-reference methodologies as well as providing a unified scheme from which we may generate approximate connected moments.
Unlike in the case of CCSD, in which we expect the typical orbital insensitivity owed to the Thouless theorem\cite{texp2}, DCM(N) does not have the orbital relaxation effects associated with singles, so meaningful energy changes may be expected when we change reference. DCM(N) also benefits from the fact that the (exact) Brueckner determinant maximizes overlap between the reference and exact wavefunctions. The convergence of the CMX series relies on that overlap and a closer starting determinant should therefore more rapidly approach the true expectation value.\cite{cioslowski1987connected2}

We can then define $\hat{H}_{\text{N}} = \hat{H} - E_{0}$ (note that all connected moments above $\mu_1$ are invariant to adding a constant to $\hat{H}$) using the expectation value of the determinant, $E_0$, as the reference result in the CMX(1) energy. It may be beneficial to discuss the nature of the low-order canonical-orbital case qualitatively before developing the general semi-canonical all-order expressions. Going further in the series, one can construct\cite{cioslowski1987molecules} 
\begin{equation}
    \mu_{2}=\braket{H_{\text{N}}^{2}}_{c} =  \braket{0|H_{\text{N}}QH_{\text{N}}|0} =\frac{1}{4}\sum\limits_{ijab}\braket{ij||ab}\braket{ab||ij}
\end{equation}
as the second-order connected moment for the canonical orbitals,  with $\braket{ij||ab}$ being the anti-symmetrized two-electron matrix elements in standard notation.\cite{szaboostland}. In particular, double substitutions from occupied orbitals $i,j,k,\cdots$ to virtual orbitals $a,b,c,\cdots$ are the only part of $Q$ that makes non-zero contributions to $\mu_2$ in Hartree-Fock. The next moment, $\mu_3$ also involves only contributions from double substitutions in the cannonical case leading to:\cite{cioslowski1987molecules}
\begin{equation}
\begin{aligned}
\mu_{3} = \braket{H_{\text{N}}^{3}}_{c} = &\sum\limits_{ijabcd}\frac{1}{8}\braket{ij||ab}\braket{ab||cd}\braket{cd||ij} + \sum\limits_{ijklab} \frac{1}{8}\braket{ij||ab}\braket{ij||kl}\braket{kl||ab}\\
&-\sum\limits_{ijkabc}\braket{ij||ab}\braket{kb||ic}\braket{kj||ac}
-\sum\limits_{ijab}\frac{1}{4}|\braket{ij||ab}|^{2}\Delta_{ij}^{ab}
\end{aligned}
\label{eq:mu3}
\end{equation}
where $\Delta_{ij}^{ab} = \epsilon_{i} + \epsilon_{j} - \epsilon_{a} - \epsilon_{b}$ are the orbital eigenenergy differences; these one-particle terms are now in the numerator of just the final term, in contrast to their position in the denominator for all terms in perturbation theory. This presents an exclusively doubles theory in the construction of the CMX(2) energy. The higher-order moments may be evaluated via the same scheme, and of course increasingly large subspaces of $Q$ begin to make contributions in the exact case. Specifically, while $H_N\ket{\Phi_0}$ contains up to double substitutions, $H_NH_N\ket{\Phi_0}$ contains single, double, triple and quadruple substitutions (S,D,T,Q). Therefore,  like the MP4 energy, $\mu_4$ also contains S,D,T,Q contributions and requires $\mathcal{O}(N^7)$ compute effort. In addition, $\mu_{4}$ contains one 'MP2-like' term with two Fock operators and three 'MP3-like' terms with a single Fock operator after combining the complex-conjugate cases.

Similar to MP5, $\mu_5$ requires $\mathcal{O}(N^8)$ computational effort to evaluate 9 classes of terms (SS, SD, ST, DD, DT, DQ, TT, TQ, QQ), plus the set of $\Delta$-containing terms, to enable construction of the CMX(3) energy via Eq. \ref{eq:CMX(3)}. Higher terms require yet greater computational effort, analogous to the corresponding M\o ller-Plesset energy terms\cite{raghavachari1990fifth,cremer2011moller}, but without orbital energy difference denominators. With the system-scaling increasing by one power per order of CMX, and the number of distinct terms or diagrams exploding, it is clear that while direct implementation of CMX(3) is challenging, CMX(4) which requires $\mu_7$, is essentially prohibitive.

Clearly there is a strong resemblance between the connected moments, $\mu_j = \braket{H^j}_c = \braket{0|H_{\text{N}}[QH_{\text{N}}]^{j}|0}$ and the corresponding $j^{\text{th}}$ order M\o ller-Plesset energy, $E^j = \braket{0|V_{\text{N}}[R_{0}V_{\text{N}}]^{j}|0}$, as each involves the same power of the fluctuation potential. There are also key differences. The most obvious distinction is the lack of the resolvent $R_{0}$ operator in $\mu_j$, which instead features a projection operator, $Q$, onto the orthogonal space. This eliminates M\o ller-Plesset energy eigenvalue denominators from the moments expressions. Another difference is that in the moments approach, there was no determination of a small parameter, unlike the perturbative case. This may help avoid perturbation theory breakdowns. When the HF reference is used, this results in moments expression where some terms include $\Delta$-dependence in the numerator, as seen in Eq. \ref{eq:mu3}. We note the advantage of additive separability of the $\Delta$-containing moment numerator terms compared to the perturbation theory energy denominators. Instead of $(j-1)$ $\Delta^{-1}$ in all terms of the $j^{\text{th}}$ order energy, there are terms containing up to $\Delta^{j-2}$ in the $j^{\text{th}}$ connected moment.

Diagonal Fock components are typically the dominant numeric value in the evaluation of $\braket{\Phi_{ij}^{ab}|H_{\text{N}}|\Phi_{kl}^{cd}}$. This fact is also why MP2 recovers most of the correlation energy in weakly correlated (large orbital gap) systems. In such systems, because the gap is large, so too is $\left| \Delta_{ij}^{ab} \right|_\infty $, and therefore the doubles amplitudes are small relative to one: ${\left|R_{0}V_{\text{N}}\ket{0}\right|}_{\infty} << 1$. Letting the largest doubles amplitude have magnitude $t_{\text{max}} << 1$, we then expect the magnitude of the MP3 energy to be smaller than MP2 by roughly this same factor because it involves one more power of $R_{0}V_{\text{N}}$. 
For MP4, the connected triples and quadruples give rise to larger eigenenergy differences such as $\Delta_{ijk}^{abc}$ and $\Delta_{ijkl}^{abcd}$ which is compounded with the presence of one more power of $R_{0}V_{\text{N}}$ in the MP4 energy. Connected doubles tend to dominate in single-reference cases which is one of the strengths of the CC methods, as its most basic form functions as a resummation of these terms through infinite order. Heuristically, the energetic contributions of the singles and triples cancels against the quadruples, contributing another advantage that has led to the success of even low-order perturbation theory and methods constrained to only doubles.

We shall now introduce the approximations to the construction of the various higher-order moments by exploiting these very same arguments, likewise considering the single-determinant case. In this instance, the dominant terms in $\mu_{3}$ ought to be the MP2-like term mentioned above. In $\mu_{4}$, the MP2-like term scaling as $|\Delta_{ij}^{ab}|^{2}$ is roughly an order of magnitude larger than the MP3-like terms scaling as $\Delta_{ij}^{ab}$, and so forth. 
Though we lack the benefit of inverse energy difference decay, we see that at each $N$-th order cumulant, the largest expected term is always contained in the connected doubles, keeping the comparison to PT relevant. Going outside the single reference case, these terms may cease to be the largest. However, unlike in M\o ller-Plesset where near degeneracies cause a singularity in the equations, the terms do not diverge in the connected moments and merely tend towards zero. This leaves the dominant small-gap terms as those containing no Fock operators. This suggests that evaluating the connected moments within just the subspace of doubles would be a worthwhile venture just as it has been in the case of CC and PT theory.

In the case of semicanonical orbitals, the exact $\mu_{2}$ includes an additional term corresponding to a sum over the Fock contributions from non-Brillouin singles, $|f_{i}^{a}|^{2}$, yet ameliorated in magnitude by their approximate Brueckner-like nature. Non-Hartree-Fock reference determinants would have these singles contributions neglected in such an approximation, which otherwise would only begin formally at $\mu_{4}$. Similar use of these orbitals in work on perturbation theory had found that approximations neglecting these singles contributions at third-order are sufficient in the construction of low-order energies\cite{Bertels20194170,Rettig20207473}. As the CMX matrix equation is of a more complex form than the sum, we desire to keep an even-tempered approach to the description of the Hilbert space at each level of the moments. Thus, we retain only doubles at all orders in this description of both the canonical and semi-canonical case of our equations.


The doubles approximation to the connected moments will be defined by replacing the resolution $(P+Q)$ by $(P+D)$ where $D$ is the doubles subspace of the orthogonal Hilbert space: $D = \frac{1}{4}\sum\limits_{ijab}\ket{\Phi_{ij}^{ab}}\bra{\Phi_{ij}^{ab}}$. In analogy to DMBPT($\infty$) and LCCD, we can begin by defining the intermediate integral tensors and now incorporating off-diagonal elements of the doubles:
\begin{equation}
    (ij||ab)_{1} = \braket{ij||ab}
\end{equation}
and
\begin{equation}
\begin{aligned}
(ij||ab)_{2} = 
&\frac{1}{4}\sum\limits_{kl}\braket{ij||kl}\braket{kl||ab} +  \frac{1}{4}\sum\limits_{cd}\braket{ij||cd}\braket{cd||ab}\\-&\sum\limits_{kc}\braket{ik||cb}\braket{jc||ka}-\sum\limits_{k}P(ij)\braket{ik||ab}f_{kj}+\sum\limits_{c}P(ab)\braket{ij||ac}f_{bc}
\end{aligned}
\end{equation}

Here $P(pq)$ is the standard anti-symmeterizer function for electrons $p$ and $q$, $(ij||ab)_{1}$ is identical to a raw anti-symmeterized two-electron integral and the terms in $(ij||ab)_{2}$ comprise the hole ladder, particle ladder, ring term, and the 'MP2-like' delta terms of $\mu_{3}$ respectively, which takes the form of $-\braket{ij||ab}\Delta_{ij}^{ab}$ in the Hartree-Fock case. From here, we may generate the 20 skeletal diagrams contributing to the $\mu_{4}$ doubles: 16 reminiscent of the doubles contribution to MP4, as well as three MP3 like terms with a single Fock contribution and one MP2 like term with two Fock contributions by a simple recursion:
\begin{equation}
\begin{aligned}
(ij||ab)_{n+1}  
& = \frac{1}{2}\sum\limits_{cd}\braket{cd||ab}(ij||cd)_{n} + 
\frac{1}{2}\sum\limits_{kl}\braket{kl||ij}(kl||ab)_{n}\\
& -\sum\limits_{k}P(ij)(ik||ab)_{n}f_{kj}
+\sum\limits_{c}P(ab)(ij||ac)_{n}f_{bc}\\
& + \sum\limits_{kc}\big[\braket{kb||jc}(ik||ca)_{n} -
\braket{jc||ka}(ik||cb)_{n} \\
& -\braket{kb||ic}(jk||ca)_{n} + 
\braket{ka||ic}(jk||cb)_{n}\big]
\end{aligned}
\end{equation}

All terms of higher order may be generated directly from the previous intermediate. Simple contraction of these intermediates results in the doubles approximation to the connected moments (DCM):

\begin{equation}
    \mu_{2n-1}^{D} = \sum\limits_{ijab}(ij||ab)_{n}(ij||ab)_{n-1} 
\end{equation}
and 
\begin{equation}
    \mu_{2n}^{D} = \sum\limits_{ijab}|(ij||ab)_{n}|^{2}
\end{equation}
Like LCCD or DMBPT($\infty$) or even CCD and CCSD, the limiting step in the formulation of the intermediates is the $\mathcal{O}(o^{2}v^{4})$ particle ladder contraction, similar to the MP3 energy or the CCD/CCSD amplitudes equation. However, unlike CCD/CCSD, this does not require solving a non-linear set of equations and, outside of the recursion, no iterations for the determination of amplitudes or energies are needed in the formation of the intermediates. One may then use the values of the approximate moments $\mu_{2n-1}^{D}$ in place of the exact $\mu_{2n-1}$ to yield the DCM(N) approximations to the full CMX(N) model. 

\section{Implementation}

The code necessary to form $(ij||ab)_{n+1}$, and therefore the necessary doubles connected moments, $\mu_{2n-1}^{D}, \mu_{2n}^{D}$ was implemented in a development version of the Q-Chem quantum chemistry program,\cite{qchem,epifanovsky2021software} which has also been used to carry out the all-electron calculations for each methodology in the following examples. Two other important aspects of the implementation should be mentioned.

First, unlike in the PT case, higher-order moments do not generally shrink in value except in the (rather pointless) case of an exact reference function, for which the second order and higher connected moments all evaluate to zero. 
However, $\mu_{k}$ has dimensionality $E^{k}$ and is dominated by terms with $(\Delta_{ij}^{ab})^{k-1}$, whose largest value is: 
\begin{equation}
  \Delta_\mathrm{max} = \max_{ijab} \Delta_{ij}^{ab}    
\end{equation}
We then redefine the energy scale such that $E'=E/\Delta_\mathrm{max}$ to keep the moments from exploding in value. In turn the scaled energy can be rescaled after the CMX algorithm is applied to obtain the final DCM(N) energy. This procedure ameliorates the round-off error introduced from operations taken between connected moments of greatly differing magnitudes.

Second, we do not explicitly invert the $(N-1)\times (N-1)$ matrix of moments that enters Eq. \ref{eq:CMX(N)}. Rather we solve a set of linear equations:
\begin{equation}
\begin{bmatrix}
\mu_{3}&\mu_{4}&\cdots&\mu_{N+1}\\
\mu_{4}&\mu_{5}&\cdots&\mu_{N+2}\\
\vdots&\vdots&\ddots&\vdots\\
\mu_{N+1}&\mu_{N+2}&\cdots&\mu_{2N-1}\\
\end{bmatrix}
\begin{bmatrix}
\zeta_{2}\\
\zeta_{3}\\
\vdots\\
\zeta_{N}
\end{bmatrix} = 
\begin{bmatrix}
\mu_{2}\\
\mu_{3}\\
\vdots\\
\mu_{N}
\end{bmatrix}
\label{eq:linear_eqs}
\end{equation}
and then evaluate:
\begin{equation}
    E^{(N)} = \mu_{1} - 
\begin{bmatrix}
\mu_{2}&\mu_{3}&\cdots&\mu_{N}
\end{bmatrix}
\begin{bmatrix}
\zeta_{2}\\
\zeta_{3}\\
\vdots\\
\zeta_{N}
\end{bmatrix}
\label{eq:CMX(N)_alt}
\end{equation}
This improves numerical stability in the case where the coefficient matrix may be near singular.

Having discussed the necessity of scaling, 
we next present some specific data for water in cc-pVDZ. 
As a reference point, the unscaled approximate moments from $\mu_{4}$ to $\mu_{39}$ span a range of over 57 orders of magnitude!
For various scaling factors, we show both the logarithm of the scaled range of the approximate DCM(N) moments and the energy differences (in $\mu$H) between calculated DCM(N) energies using different rescalings in Table \ref{Table:H2O-scaling}. 

\begin{table}
\begin{center}
\begin{tabular}{ |c|c|c|c|c|c|c|c|c| } 
 \hline\hline
 \multirow{2}{*}{Method} & \multicolumn{8}{c|}{Scaling Factor} \\ 
 \cline{2-9}
  & 0.85 & 0.90 & 0.95 & 1.00 & 1.05 &  1.15 & 1.20 &  1.25   \\
 \hline
DCM(14)	&	119.5	&	127.9	&	-6.0	&	-6.0	&	-4.4	&	-0.4	&	0.8	&	1.0	\\
DCM(15)	&	138.6	&	-12.5	&	-8.8	&	-6.8	&	-3.6	&	2.7	&	5.7	&	10.1	\\
DCM(16)	&	-25.2	&	-20.5	&	-15.6	&	-10.7	&	-5.5	&	5.6	&	12.0	&	18.1	\\
DCM(17)	&	35.0	&	42.2	&	49.9	&	58.1	&	-3.2	&	1.7	&	4.9	&	9.0	\\
DCM(18)	&	38.2	&	49.1	&	-11.8	&	-8.3	&	-4.6	&	2.7	&	7.3	&	10.5	\\
DCM(19)	&	47.1	&	63.4	&	-13.2	&	-9.3	&	-5.0	&	4.5	&	9.0	&	13.8	\\
DCM(20)	&	59.8	&	-20.6	&	-15.4	&	-10.7	&	-5.3	&	5.6	&	11.1	&	15.5	\\
\hline
Log Range	&	3.66	&	2.80	&	1.97	&	1.19	&	0.52	&	1.18	&	1.78	&	2.36	\\

 \hline\hline
 \multicolumn{6}{c}
\small\text{Logarithmic Range of 1.1 is 0.57}
\end{tabular}
\end{center}
\caption{Difference in DCM(N) ($\mathrm{N}=14\cdots20$) energies for water  (in $\mu$H) and logarithmic range of calculated doubles moments as a function of scaling factor on top of  $\Delta_\mathrm{max}^{-1}$. Energies are relative to 1.1}
\label{Table:H2O-scaling}
\end{table}

The range of magnitudes of the scaled connected moments is very satisfactory ($< 10^4$) for all scalings considered. We focus on the behavior of the the highest-order DCM(N) energies for different scaling factors multiplied by $\Delta_\mathrm{max}^{-1}$, from 0.85 to 1.25 and compared against the value of 1.1. Even including the unscaled approach, all energies through DCM(5) are the same within machine precision, while this is true of all scaled moment approaches tested here through DCM(11). The fact that the higher order DCM(N) energies show differences of $\mathcal{O}(10^0-10^2)$ $\mu$H suggests that solving the linear equations, Eqs. \ref{eq:linear_eqs} may involve some ill-conditioning even in scaled energy units.

\begin{table}
\begin{center}
\begin{tabular}{ |c|c|c|c|c| } 
 \hline\hline
 \multirow{2}{*}{Method} & \multicolumn{2}{c|}{H$_{2}$O} & \multicolumn{2}{c|}{F$_{2}$} \\
 \cline{2-5}
  & HF & OO & HF & OO   \\
  \hline
DCM(6)	&	5.66	&	5.66	&	5.88	&	5.89	\\
DCM(9)	&	10.28	&	10.28	&	10.41	&	10.40	\\
DCM(10)	&	12.34	&	12.34	&	12.39	&	12.39	\\
DCM(11)	&	13.56	&	13.55	&	13.53	&	13.52	\\
DCM(12)	&	15.07	&	15.18	&	14.98	&	14.96	\\
DCM(13)	&	15.65	&	16.39	&	15.56	&	15.51	\\
DCM(14)	&	16.01	&	16.78	&	15.45	&	15.74	\\
DCM(15)	&	15.96	&	16.82	&	15.47	&	15.77	\\
DCM(16)	&	15.86	&	16.14	&	16.33	&	15.64	\\
DCM(17)	&	17.67	&	16.21	&	16.04	&	15.68	\\
DCM(18)	&	15.86	&	16.14	&	16.33	&	15.64	\\
DCM(19)	&	16.29	&	16.47	&	16.16	&	15.49	\\
DCM(20)	&	16.13	&	16.73	&	16.69	&	15.89	\\

 \hline\hline
\end{tabular}
\end{center}
\caption{Logarithm of the condition number of the coefficient matrix in Eq. \ref{eq:linear_eqs}, for various DCM(N) and oo:DCM(N) orders of H$_{2}$O and F$_{2}$}
\label{Table:log-scaling}
\end{table}

To assess whether or not  Eqs. \ref{eq:linear_eqs} can be ill-conditioned, Table \ref{Table:log-scaling} shows the logarithm of the condition numbers for both the H$_{2}$O and F$_{2}$ systems. Quite plainly for higher orders of DCM(N) or oo:DCM(N), the condition number becomes large enough that the maximum forward-backward error can in theory make the results meaningless in double precision (64 bit) arithmetic. For instance by DCM(13), the condition number is larger than the inverse of 64 bit machine precision. Despite this, the error in-practice is the quantity of interest and many linear algebra techniques are widely employed in the context of poorly-conditioned matrices. Table \ref{Table:LinAlg_error} contains errors of DCM(N) from the exact FCI for the fluorine dimer for various linear algebra approaches when fed moments evaluated in double precision. The only difference between trials are small perturbations related to values beyond arithmetic precision. By comparison, the errors associated with CCSD and CCSD(T) for this example are 11.14 mH and 1.86 mH, respectively. As will be examined in more detail later, fluorine has the largest fluctuations between DCM(N) orders we have observed.  As can be seen in As is evident from Table \ref{Table:LinAlg_error}, the DCM(N) errors exhibit a sawtooth pattern with trough-to-peaks near 0.8mH when utilizing Armadillo done with a Moore-Penrose psuedo-inverse (xGELSD) algorithm\cite{sanderson2016armadillo}, a common fallback strategy for poorly-conditioned matrices. This may be compared to other approaches, such as the standard xGESV class of subroutines in LAPACK\cite{lapack99} or the NumPy\cite{harris2020array} routines likewise based on the xGEEV subroutines. 
\begin{table}
\begin{center}
\begin{tabular}{ |c|c|c|c|c|c|c| } 
 \hline\hline
 \multirow{2}{*}{Trial} & \multicolumn{6}{c|}{Method}\\
 \cline{2-7} &
 DCM(11)	&	DCM(13)	&	DCM(15)	&	DCM(17)	&	DCM(19)	&	DCM(20)	\\
\hline
\multicolumn{7}{|c|}{Moore-Penrose Psuedo-Inverse (xGELSD)} \\
\hline
1	&	5.608	&	4.844	&	5.625	&	5.028	&	5.568	&	5.747	\\
2	&	5.613	&	4.873	&	5.623	&	5.036	&	5.568	&	5.746	\\
3	&	5.611	&	4.84	&	5.622	&	5.026	&	5.567	&	5.747	\\
4	&	5.601	&	4.845	&	5.623	&	5.026	&	5.568	&	5.748	\\
\hline
\multicolumn{7}{|c|}{Standard LAPACK (xGESV)} \\
\hline
1	&	5.608	&	4.839	&	4.762	&	4.795	&	3.698	&	3.156	\\
2	&	5.614	&	4.863	&	5.036	&	4.760	&	3.319	&	3.254	\\
3	&	5.611	&	4.835	&	4.911	&	4.784	&	2.92	&	3.437	\\
4	&	5.601	&	4.826	&	4.794	&	4.799	&	3.428	&	3.810	\\
\hline
\multicolumn{7}{|c|}{Standard NumPy (xGEEV)} \\
\hline
1	&	5.610	&	4.775	&	4.704	&	4.832	&	4.441	&	4.375	\\
2	&	5.614	&	4.862	&	5.047	&	4.727	&	4.402	&	4.272	\\
3	&	5.613	&	4.755	&	4.844	&	4.787	&	4.22	&	4.477	\\
4	&	5.600	&	4.910	&	4.827	&	4.85	&	3.574	&	4.367	\\

 \hline\hline
\end{tabular}
\end{center}
\caption{Error from the exact energy (mH) for \ce{F2} for various DCM(N) methods using 3 different linear algebra packages to solve Eqs. \ref{eq:linear_eqs}. Trial 1 corresponds to use of moments as calculated in double precision arithmetic, while trials 2-4 involve a random perturbation added to the calculated moments at the level of $10^{-13}$.}
\label{Table:LinAlg_error}
\end{table}

With both of the standard algorithms not incorporating techniques to manage singular values, we see a strong inconsistency between trials. While in all cases the low-order DCM(11) sees errors near our 10 $\mu$H criteria, by DCM(13) the non-adapted codes begin to see maximum differences between runs up to the quite striking differences of 0.78 and 0.87 mH for DCM(19). By contrast the Moore-Penrose algorithm used here ameliorates this difference to 1 $\mu$H, while its largest difference between runs is in DCM(13) at 33 $\mu$H. Unfortunately, the nature of the SVD approximation results in a dropping quality of its spectrum, which becomes noticeable for some examples in the test set. After DCM(13), which are fairly similar energetically for all approaches, the average of possible trials for the standard algorithms would be closer to the exact than the SVD by ~1-2 mH. While this is significant, stronger consistency between runs is always a desirable trait to balance against the improvement. In either event, every single run presented is well below the CCSD energetic error of 11.14 mH, an encouraging sign despite the forward-backward error of the moment matrix. Finally, in passing, we note that precision errors introduced by the operations within the linear algebra itself have also been considered, but were ultimately significantly smaller than both of the other errors presented here at between 0.1 and 100 nH.

Based on the data presented, as well as tests on other members of the G1 data set, we decided to employ a scaling factor of ${1.1}\Delta_\mathrm{max}^{-1}$ in the moment generating procedure, followed by multiplying its reciprocal value back to the DCM(N) energy in order to re-scale back to atomic units, as well as the use of the Moore-Penrose SVD through the Armadillo C++ Library utilizing xGELSD. 

\section{Results and Discussion}

\subsection{1. G1 Test Set}

We shall address the accuracy of the DCM method for a wide range of small molecules as a measure of its usefulness for the description of chemical systems. Towards this end, we compare the energies of CMX(2), DCM(N), MP2, CCD, CCSD, and CCSD(T) against ASCI+PT2 with a cc-pVDZ basis across 55 systems contained in the first-row G1 test set\cite{Tubman2018ModernAT,g1_1,g1_2, g1_3} at the geometries described in the original paper and its references. We note that we use the original geometries of the G1 test set in place of the Feller geometries, with a re-evaluated ASCI+PT2 for the exact energies. All ASCI+PT2 calculations were converged to within 10 $\mu$H accuracy and serve as an effectively converged full CI calculation. 

As the nature of the DCM series' convergence was not known before this work, we will report its behavior for specific systems in addition to the general statistics of the test set. In particular, low-order CMX has been reported to occasionally converge to excited states in the case of a dominant excited state configuration.\cite{cmx_excite} Erroneously, CMX was thought to converge to incorrect values in seminal works\cite{cmx2}, but it was demonstrated that this was an excited state energy of the system\cite{Amore_2011}). In the spectral representation of Horn-Weinstein, as one reaches high orders, all excited states ought to be sufficiently damped away such that only the ground state survives. However this condition is not necessarily ensured in our subsection of the Hilbert space limited to doubles, nor is the monotonic nature of the series. It is  a useful task to assess the convergence of the DCM(N) series to determine if the Nth term in the series improves over the previous value. We begin by inspecting some representative systems of DCM(N)'s behavior in the G1 test set: H$_{2}$O, F$_2$, C$_{2}$H$_{6}$, CN, and Li$_2$.

\subsection{A. H$_{2}$O and Model Scaling}

\begin{figure}
\centering
\includegraphics[width=12cm]{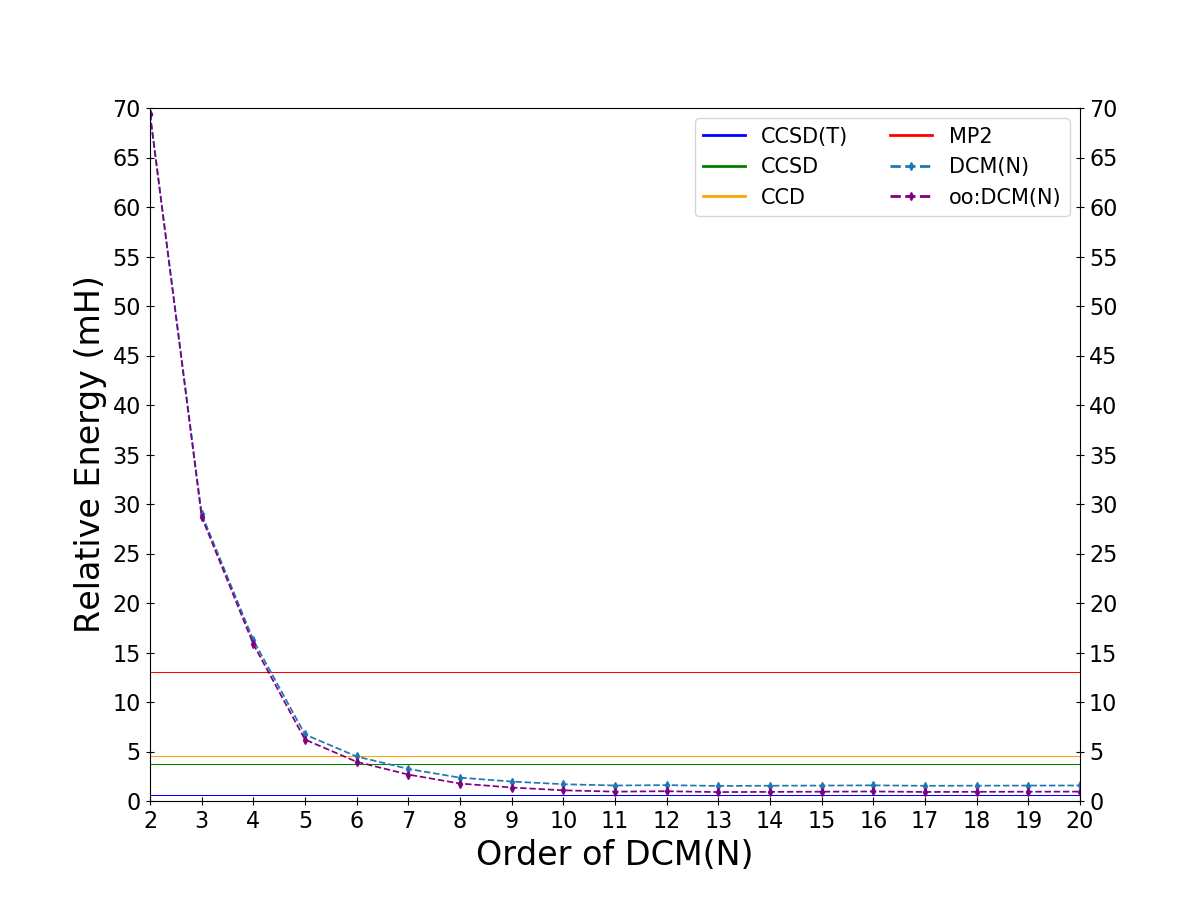}
\caption{The total energy of water in cc-pVDZ for various DCM(N) and oo:DCM(N) orders.}
\label{fig:h2o_dcm}
\end{figure}

\begin{figure}
\centering
\includegraphics[width=12cm]{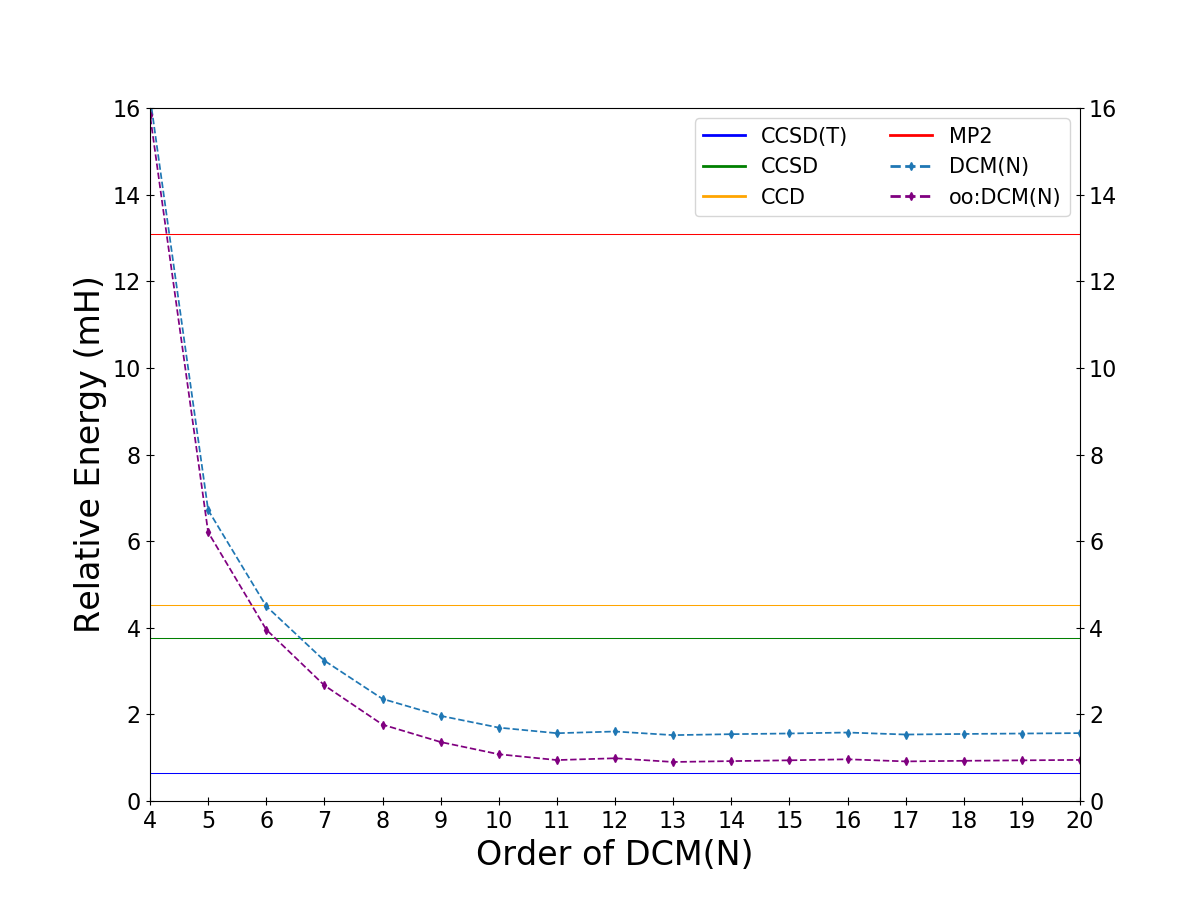}
\caption{The total energy of water in cc-pVDZ for various DCM(4) and oo:DCM(4) and beyond.} 
\label{fig:h2o_zoom}
\end{figure}

Water has served as a test molecule for the behavior of high-order perturbation theory since early algorithms were developed, including sets of selective summations towards the infinite limit which we seek to likewise address\cite{bartlett1977comparison}. Therefore, it serves as a useful comparison to begin to understand the behavior of the DCM(N) series. In Figure \ref{fig:h2o_dcm}, we depict order-by-order the DCM(N) and oo:DCM(N) energies relative to the exact energy of H$_{2}$O in the cc-pVDZ basis, while a more resolved image focused on just the 4th order and higher terms are depicted in Figure \ref{fig:h2o_zoom}. The energy errors of some common methods (MP2, CCD, CCSD and CCSD(T)) relative to the exact energy are depicted as horizontal lines. 

Referring to Fig. \ref{fig:h2o_dcm}, CMX(2) and oo:DCM(2) are evidently a very poor description of electron correlation, with over 5 times larger error than the (cheaper) MP2 method. We also see that low-order DCM(N) is insufficient for quantitative accuracy; DCM(N) is inferior to CCD and CCSD until DCM(6) and DCM(7), respectively. However the fact that the DCM(N) sequence crosses over with CCD and CCSD is exciting, and, perhaps, unexpected. Fig. \ref{fig:h2o_zoom} provides a zoomed in view of the approach of the sequence of DCM(N) values towards a potential DCM($\infty$) limit. The sequence settles down to be close to an apparent limit by DCM(11). There is non-monotonic behavior with variations in the value differing by less than 0.1 mH until DCM(14); after this all energetic changes are no larger than 30 $\mu$H.

This convergence behavior is reminiscent of the best behavior seen in the case of conventional M\o ller-Plesset perturbative series, such as for water at $R_e$ in very small basis sets.\cite{handy1985convergence} However, even well-behaved closed shell systems such as the Ne atom exhibit poor convergence and even divergence in slightly larger basis sets such as aug-cc-pVDZ.\cite{christiansen1996inherent} 
By contrast, the exact CMX series has a monotonic approach to exactness. However, we have no guarantee that our doubles approximation will replicate the monotonicity of CMX, 
and therefore oscillations are a possibility. 
In this regard, the data shown in Figs. \ref{fig:h2o_dcm} and \ref{fig:h2o_zoom} is very encouraging as the DCM(N) sequence appears quite stable, at least up through DCM(20) in the tested cc-pVDZ basis. One can explain this by the fact that the products of small eigenenergies in the denominator contributing to this behavior in perturbation theory go towards zero in the corresponding moments. Most intriguingly, both the HF- and oo- based DCM(N) methods outperform their fellow O(N$^6$) CCD and CCSD methods in water. The DCM(N) methods even recover a substantial fraction of the triples contribution, particularly when using the OOMP2 reference determinant.

\subsection{B. CN \& C$_{2}$H$_{6}$}

\begin{figure}
\centering
\includegraphics[width=12cm]{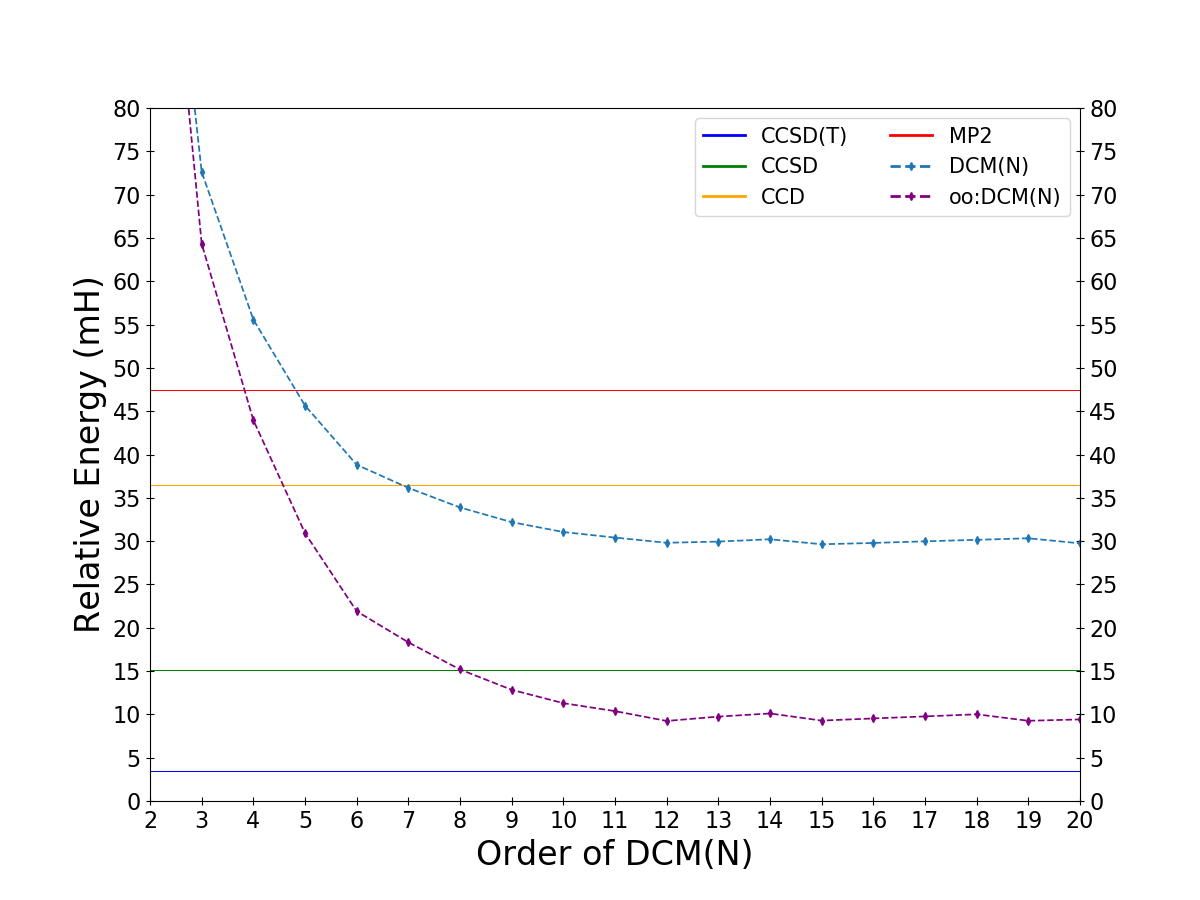}
\caption{The total energy of the CN radical in cc-pVDZ for various methods by order of DCM(N) and oo:DCM(N).} 
\label{fig:cn_energy}
\end{figure}

The cyano radical and ethane present two interesting examples for the behavior of the DCM(N) sequence of energies. They are, respectively, the systems in which DCM(N) perform the worst and best compared to CCSD relative to the exact energy. In the case of CN, CCSD outcompetes high-order DCM(N) by about 15 mH, with Figure \ref{fig:cn_energy} showing the values of DCM(N) by order, while for ethane high-order DCM(N) outcompetes by about 7.6 mH.

In fact, CN exhibits the largest error across the entire test set for DCM(N). We see an immediate clue as to the origin of the error from the fact that CCD shows an even larger error, that is about 20\% worse than the DCM(N) sequence, and over twice as large as CCSD. If this is truly an issue associated with the lack of singles, approximate Brueckner-like orbitals would help ameliorate the error associated with orbital relaxation. Indeed the much better performance of oo:DCM(N) suggests this is the case, as the error is reduced by roughly a factor of 3 (or 20 mH), with the correlation energy below that of CCSD by DCM(9). As a result, CN is not the worst performer within the dataset for oo:DCM(N). This title instead belongs to SO$_2$, in which both oo:DCM(13) and oo:DCM(20) have a 31.6 mH error compared to CCSD's 25.1 mH. 

\begin{figure}
\centering
\includegraphics[width=12cm]{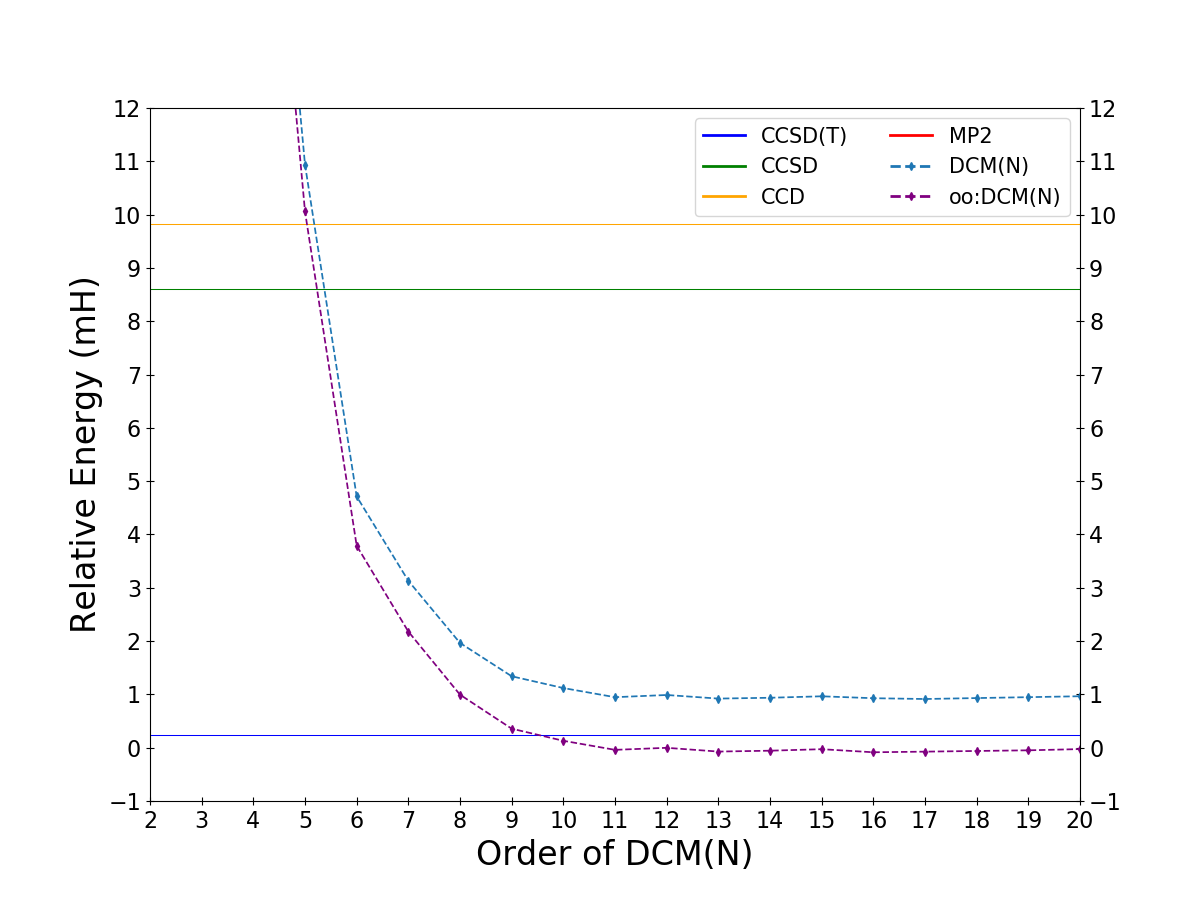}
\caption{The total energy of the ethane molecule in cc-pVDZ for various methods by order of DCM(N) and oo:DCM(N).} 
\label{fig:ethane_energy}
\end{figure}

The ethane molecule presents an interesting contrast in that CCSD is most out-performed by the DCM(N) sequence. 
Figure \ref{fig:ethane_energy} shows the behavior of the energy with respect to DCM(N) order as before, with CCSD and CCD showing errors of roughly 8 and 9 mH relative to the CCSD(T) level of chemical accuracy. 
This example is not so much a case of poor performance of either CCD or CCSD, as remarkably good performance of the DCM(N) sequence. The limiting error of DCM(N) is only about 1 mH, and the oo:DCM(N) sequence has error that is less than 0.1 mH, and is clearly superior to CCSD(T), which is in error by 0.24 mH. This example suggests that the importance of triple substitutions in the oo:DCM(N) hierarchy may be less than in the usual coupled cluster hierarchy.

\subsection{C. F$_{2}$ \& Li$_{2}$}


\begin{figure}
\centering
\includegraphics[width=12cm]{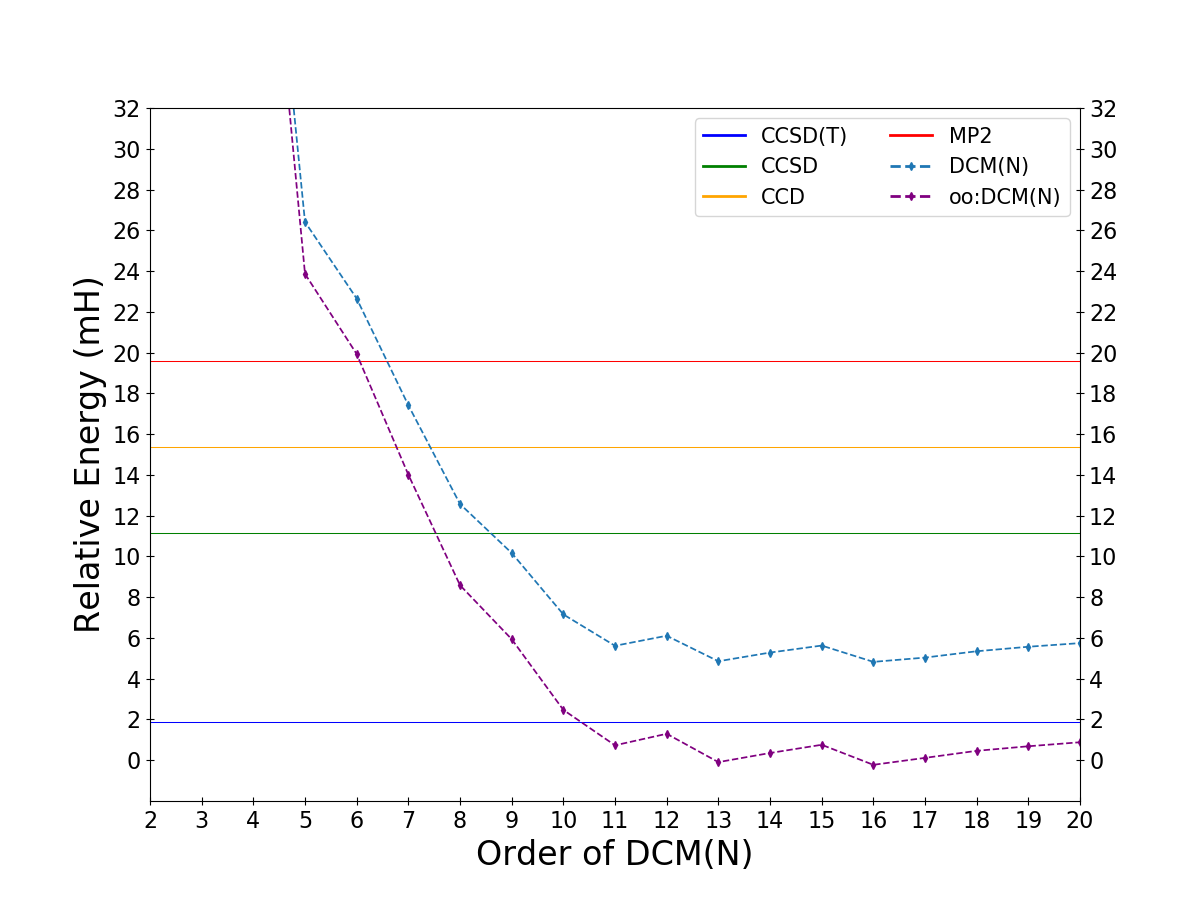}
\caption{The total energy of the F2 molecule in cc-pVDZ for various methods by order of DCM(N) and oo:DCM(N).} 
\label{f2_energy}
\end{figure}

Figure \ref{f2_energy} shows the correlation energy errors of the DCM(N) and oo:DCM(N) sequences relative to several standard methods for the case of the fluorine dimer using restricted orbitals. \ce{F2} is known to exhibit significant diradicaloid character, and in fact is not even bound at the mean-field Hartree-Fock level. In this case, we see the most significant oscillatory patterns in the DCM(N) energies out of the full set of molecules. Referring back to the investigation of the forward-backward error, we presume that a significant portion of this is due to the choice of SVD algorithm, rather than a true oscilation between the series given infinite precision. This may be elucidated by performing the calculation with increased precision in the moments, as even one or two additional digits would drop the magnitude to within chemical significance for the most severe example we have encountered thus far. Non-smooth decreases in the correlation energy occur until DCM(11), at which point the more minor changes we saw before swell to span a range of 1.5 mH between the troughs and peaks of oo:DCM(12)/oo:DCM(13) and oo:DCM(16)/oo:DCM(17), with slightly smaller values in the Hartree-Fock reference case. Despite this slightly troubling behavior, the DCM(N) values significantly surpass the CCSD energy even before the oscillations occur. In the oo:DCM(N) case, the energies even improve upon the costlier CCSD(T) method. 

\begin{figure}
\centering
\includegraphics[width=12cm]{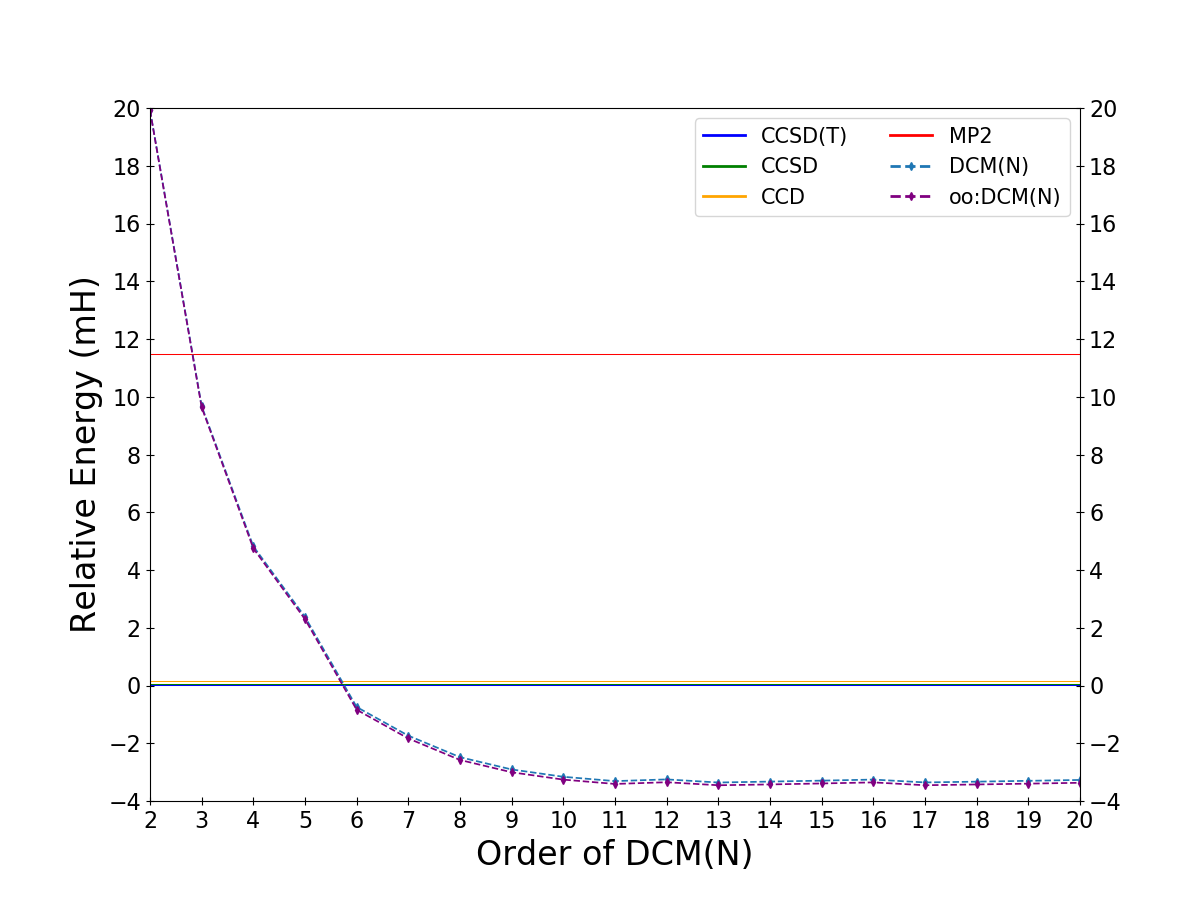}
\caption{The total energy of \ce{Li2} in cc-pVDZ for various methods by order of DCM(N) and oo:DCM(N).} 
\label{fig:li2_energy}
\end{figure}

The lithium dimer exhibits nonvariational behavior for the DCM(N) energies, as shown in Figure \ref{fig:li2_energy}. While CCD, CCSD, and CCSD(T) are all within chemical accuracy for this effectively 2-electron system, both the DCM(N) and oo:DCM(N) sequences show more than 10 times larger error, with energies going below the exact energy by up to 3.4 mH. 
The slightly worse behavior in the case of oo:DCM(N) suggests that this is not a deficiency associated with the singles. Whether the resummation structure or the moment structure is the cause of this undesirable behavior is unclear to us at present. 
Non-variational behavior is observed in LiH also, although the effect is less severe.
We note our presentation of the order-by-order analysis for various molecules illustrates that different behavior can arise from one system to the next. We additionally note that it seems to be generally true that 
the first non-monotonic values occur between DCM(11) and DCM(14), which is also the range where values appear to also approach limiting values in well-behaved cases.

\subsection{D. Results across the G1 Test Set}

The statistical results across the G1 test set are contained within Table \ref{Table2}, where the RMSD of the various methodologies discussed are shown for both the Hartree-Fock and OO reference determinants. The data has been arranged with respect to the entire set as well as two subsets: one containing molecules with only row 1 and 2 atoms, followed by the complement subset containing row 3 atoms. 
As expected, low-order DCM(N) energies are not useful methods, in line with our understanding of the poor energetic performance of CMX from past investigations. We also note there is no improvement, and in fact even slightly poorer performance of oo:DCM(2) and oo:DCM(3) compared to their Hartree-Fock counterparts.

\begin{table}
\begin{center}
\begin{tabular}{ |c|c|c|c|c|c|c|c| } 
 \hline\hline
 \multirow{2}{*}{Method} & \multicolumn{3}{c|}{HF RMSD} &\multirow{2}{*}{Method} & \multicolumn{3}{c|}{OO RMSD} \\ 
 \cline{2-4} \cline{6-8}
  & All & Row 1+2 & Row 3 & & All & Row 1+2 & Row 3   \\
 \hline
CCSD(T)	&	1.42	&	1.32	&	1.53	&	-	&	-	&	-	&	-	\\
CCSD	&	9.98	&	9.49	&	10.58	&	-	&	-	&	-	&	-	\\
CCD	&	13.60	&	13.58	&	13.62	&	-	&	-	&	-	&	-	\\
MP2	&	27.80	&	24.93	&	31.12	&	-	& - & -	& - 	\\
CMX(2)	&	114.29	&	101.78	&	128.67	&	oo:DCM(2)	&	115.29	&	101.72	&	130.76	\\
DCM(3)	&	76.69	&	48.14	&	102.39	&	oo:DCM(3)	&	77.04	&	46.79	&	103.79	\\
DCM(4)	&	55.60	&	30.96	&	76.45	&	oo:DCM(4)	&	55.34	&	28.93	&	77.05	\\
DCM(5)	&	30.22	&	18.89	&	40.40	&	oo:DCM(5)	&	29.32	&	16.07	&	40.45	\\
DCM(6)	&	23.44	&	14.56	&	31.40	&	oo:DCM(6)	&	22.20	&	11.36	&	31.02	\\
DCM(7)	&	19.88	&	12.26	&	26.67	&	oo:DCM(7)	&	18.50	&	8.76	&	26.18	\\
DCM(8)	&	14.90	&	10.60	&	19.07	&	oo:DCM(8)	&	13.07	&	6.86	&	18.18	\\
DCM(9)	&	12.46	&	9.59	&	15.40	&	oo:DCM(9)	&	10.43	&	5.73	&	14.38	\\
DCM(10)	&	11.26	&	8.90	&	13.73	&	oo:DCM(10)	&	9.16	&	4.96	&	12.67	\\
DCM(11)	&	10.45	&	8.51	&	12.53	&	oo:DCM(11)	&	8.22	&	4.54	&	11.33	\\
DCM(12)	&	10.31	&	8.44	&	12.32	&	oo:DCM(12)	&	8.10	&	4.50	&	11.15	\\
DCM(13)	&	10.43	&	8.32	&	12.64	&	oo:DCM(13)	&	8.27	&	4.35	&	11.50	\\
DCM(14)	&	10.27	&	8.44	&	12.23	&	oo:DCM(14)	&	8.06	&	4.47	&	11.09	\\
DCM(15)	&	10.33	&	8.38	&	12.41	&	oo:DCM(15)	&	8.11	&	4.39	&	11.22	\\
DCM(16)	&	10.36	&	8.24	&	12.58	&	oo:DCM(16)	&	8.19	&	4.27	&	11.40	\\
DCM(17)	&	10.41	&	8.32	&	12.61	&	oo:DCM(17)	&	8.22	&	4.35	&	11.41	\\
DCM(18)	&	10.37	&	8.40	&	12.45	&	oo:DCM(18)	&	8.14	&	4.43	&	11.24	\\
DCM(19)	&	10.38	&	8.43	&	12.44	&	oo:DCM(19)	&	8.15	&	4.39	&	11.28	\\
DCM(20)	&	10.34	&	8.27	&	12.51	&	oo:DCM(20)	&	8.15	&	4.28	&	11.34	\\

 \hline\hline
 \multicolumn{5}{c}
     \small $^{ }$
\end{tabular}
\end{center}
\caption{RMSD of approximate methods for the correlation energy relative to (nearly) exact selected configuration interaction results (in mH) across the G1 test set, evaluated in the cc-pVDZ basis.}
\label{Table2}
\end{table}

Consistent with our examination of individual cases, the most exciting results concern the convergence of DCM(N) energies, and especially the oo:DCM(N) energies to very useful values by DCM(11-14) or oo:DCM(11-14). Seeking a direct comparison of the purely doubles methods with similar orbitals, we see a notable improvement over the CCD energies as one iterates through the DCM(N) cycles across both the whole set and the partitions. This improvement is most significant for row 1 and 2 molecules, in which all orders from 12th and onwards improve upon CCD by a margin of over 5 mH. The improvement is less notable among molecules with row 3 atoms, where the same comparison only yields a single mH improvement. Together, this results in an overall improvement of 3.3 mH in the energies of the double methods, despite the lack of the disconnected-cluster contributions allowing for indirect excitations related to $\hat{T}_{2}^{2}$ in the CCD amplitudes.

Moving to the comparison of CCSD, it is useful to use the oo:DCM(N) values not just for their quantitative improvement, but also due in part to their approximate Brueckner-like nature that ameliorates orbital relation effects, a property already inherent to CCSD with its orbital insensitivity. For the row 1 and 2 containing molecules, oo:DCM(N) of 12th order and higher improves the energies by a margin of over 5 mH, while across the entire set this value is a more modest value of just under 2mH due to the smaller improvements in the row 3 molecules. Whether the smaller improvements for row 3 containing molecules are a feature of the summation scheme or the moments approach more generally is not clear without additional investigation.  It is worth noting that values of oo:DCM(N) after 12th-order are superior relative to CCSD for 49 of the 55 molecules in the G1 test set. Most of the RMSD error stems from a few systems in which the errors are high, such as the previously mentioned SO. 
In fact, the successful cases are significant enough for oo:DCM(N) to outcompete even CCSD(T) in 15 of the systems tested here. 

We also note that the same trend of rising values found in the order-by-order DCM(N) and oo:DCM(N) energy analysis for individual cases holds true in a statistical sense as well. There is some variation in the solutions, but DCM(14) exhibits the smallest RMSD in both the HF and OO cases as energies begin to oscillate. Despite this, anything 12th-order and higher are no more than 0.21 mH above this energy error for both sets of reference determinants.

As mentioned above, DCM(N) scales as the same polynomial power of system size as the CCD and CCSD equations. Specifically, the construction of $(ij||ab)_{n}$ is needed to assemble $\mu_{2N-1}^{D}$, and making $(ij||ab)_{n}$ involves $N-1$ particle ladder contractions, which roughly makes DCM(N) as costly as $N-1$ iterations of the CCD amplitudes. 
While the pilot code has not undergone heavy optimization, it may still be useful to mention preliminary timings in constructing the amplitudes versus the doubles connected moments. For the largest system, Si$_2$H$_6$, CCSD yields 81.3 seconds per cycle, for a total of 650.6 seconds in solving the Coupled Cluster equations. By comparison, DCM(N) required 42.6 seconds per cycle, resulting in DCM(14) being just under the CCSD timing at 596.5 seconds. 
We can anticipate that when CCSD convergence is slow, the DCM(N) models are a fairly straightforward way to evaluate correlation energies with comparable, or even improved, values due to its single shot nature.

\subsection{B. N$_{2}$ Dissociation}

The potential energy surface corresponding to breaking the triple bond of \ce{N2} has served as a benchmark case in the evaluation of new correlation methodologies such as CI and CC studies through very high orders, and new multi-reference CC approaches, including the many different flavors of Fock-Space MRCC, State-Universal MRCC, and Valence-Universal MRCC\cite{KROGH2001578, n2_garnet, mr_n2}.
\ce{N2} has several valuable properties for benchmarking, beginning with the need to treat both dynamic and static correlation along the dissociation coordinate. While fairly well described by a single determinant at the equilibrium geometry, the triple bond dissociation requires the interaction of 6 active electrons to recouple the two $^4$N atoms. \ce{N2} is also an excellent case for testing spatial and spin symmetry (breaking) as a function of bond-breaking. While low-order CC techniques fail qualitatively when using a spin-restricted reference determinant, the use of a spin-polarized reference allows one to obtain quantitative accuracy at dissociation at the cost of losing the spin purity inherent to the true wavefunction. 

\begin{figure}
\centering
\includegraphics[width=12cm]{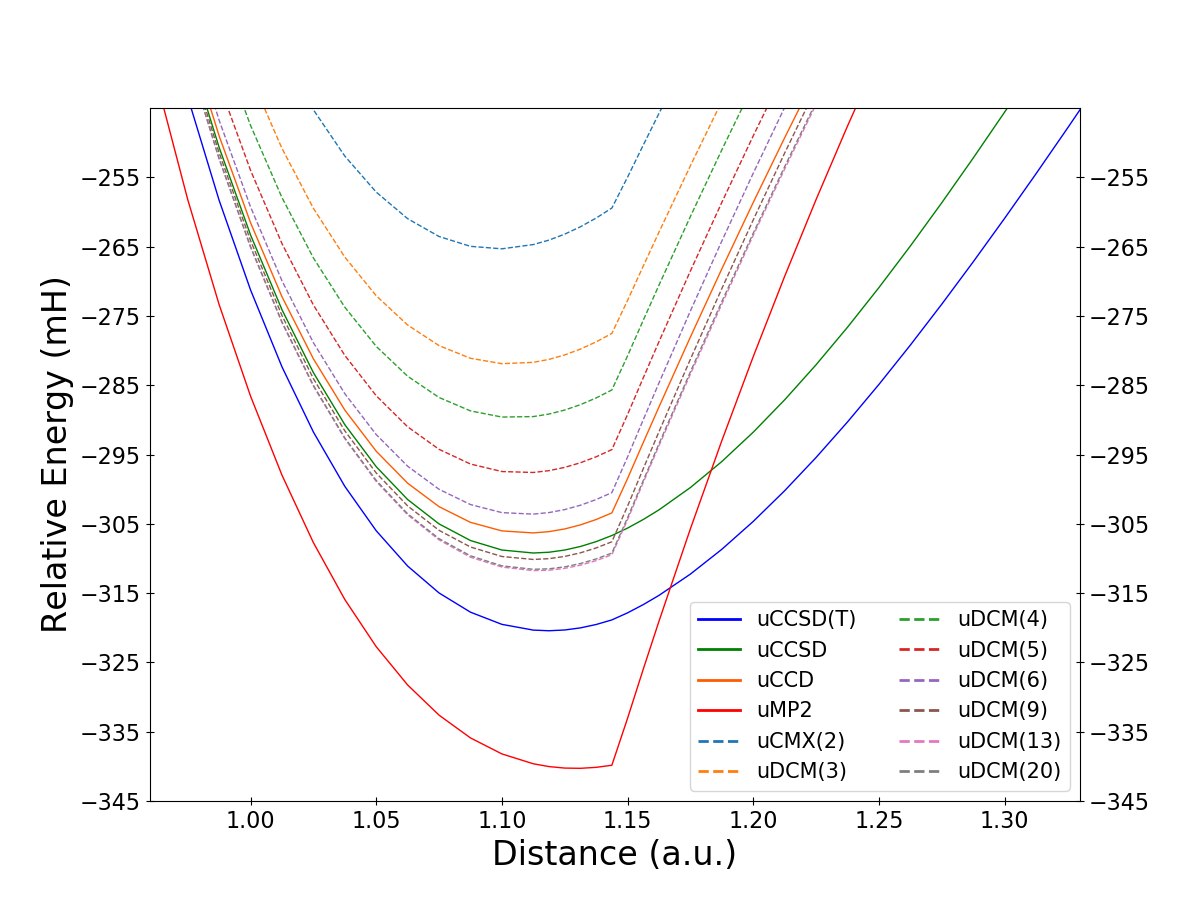}
\caption{Approximations to the total energy of N$_{2}$ in the vicinity of the equilibrium geometry with a UHF reference for correlation methods including representatives of the uDCM(N) sequence. Energies are reported relative to two $^4$N atoms for each method.}
\label{fig:n2_uhf}
\end{figure}

Figure \ref{fig:n2_uhf} shows \ce{N2} dissociation in the cc-pVDZ basis using a spin-polarized reference determinant, comparing UMP2, UCCD, UCCSD, and UCCSD(T) against the DCM(N) sequence. To begin, we focus on the convergence of the sequence near the bottom of the well for the Hartree-Fock case before considering the entire curve or the oo:DCM(N) case. For readability, we have omitted some orders, but nevertheless one can see the same trends as before: there is variation of the DCM(N) energies on the sub-mH energy scale after DCM(11). 
As usual, the traditional purely double methods of uMP2 and uCCD both exhibit a very visible first derivative discontinuity\cite{kurlancheek2009violations} at the Coulson-Fischer point, while uCCSD and uCCSD(T) are both able to circumvent this behavior (at least on the graphical scale) via the inclusion of singles. Lacking orbital relaxation due to singles, the uDCM(N) family of methods share the behavior of uMP2 and uCCD with a pronounced first derivative kink at all orders. However, the actual energetic predictions of the uDCM(N) sequence follows the trends as seen for the G1 test set: uDCM(N) surpasses uCCD by uDCM(6) regardless of whether we are in the unpolarized or spin-polarized regimes.

\begin{figure}
\centering
\includegraphics[width=12cm]{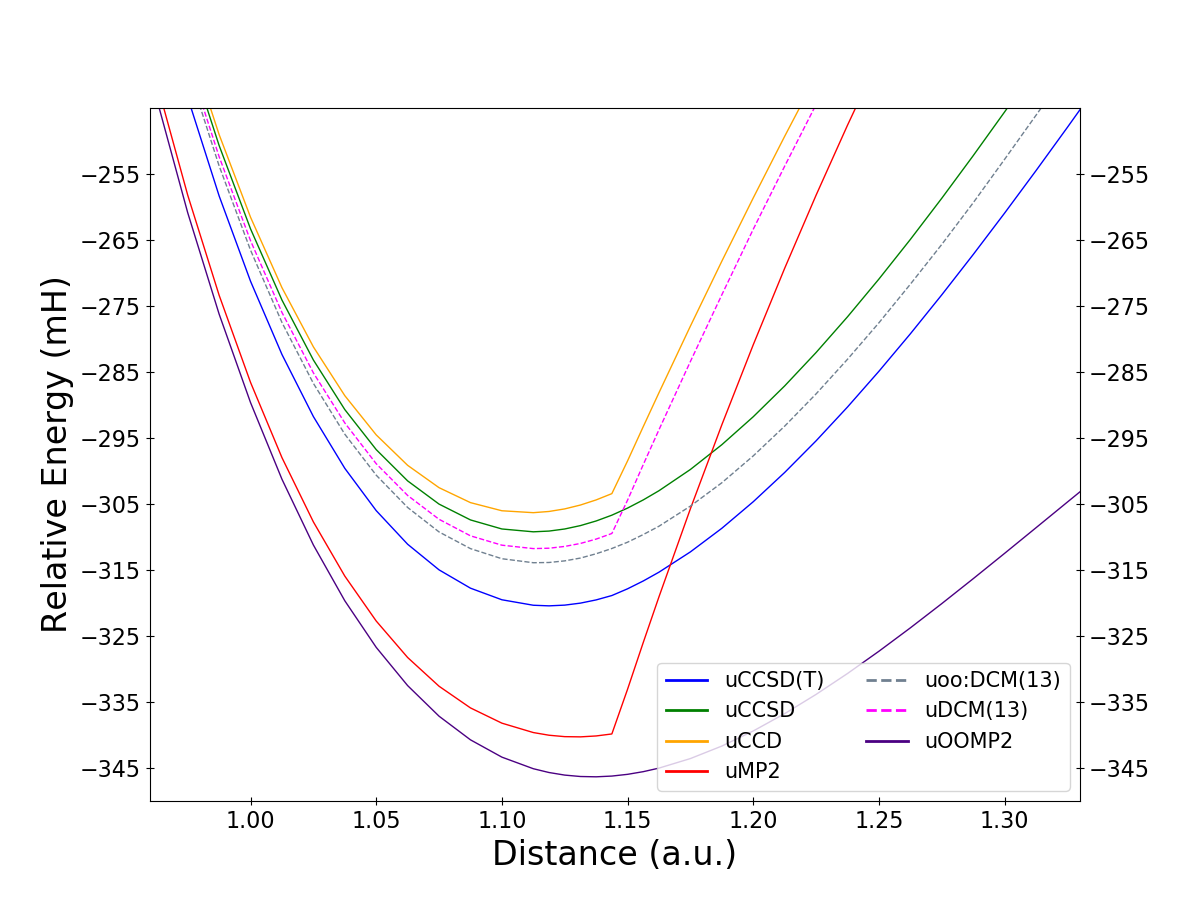}
\caption{Approximations to the total energy of N$_{2}$ in the vicinity of the equilibrium geometry with a UHF reference for correlation methods including uDCM(13) and uooDCM(13) as high order representatives of the uDCM(N) sequence. Energies are reported relative to two $^4$N atoms for each method.}
\label{fig:zoom_oon2}
\end{figure}

\begin{figure}
\centering
\includegraphics[width=12cm]{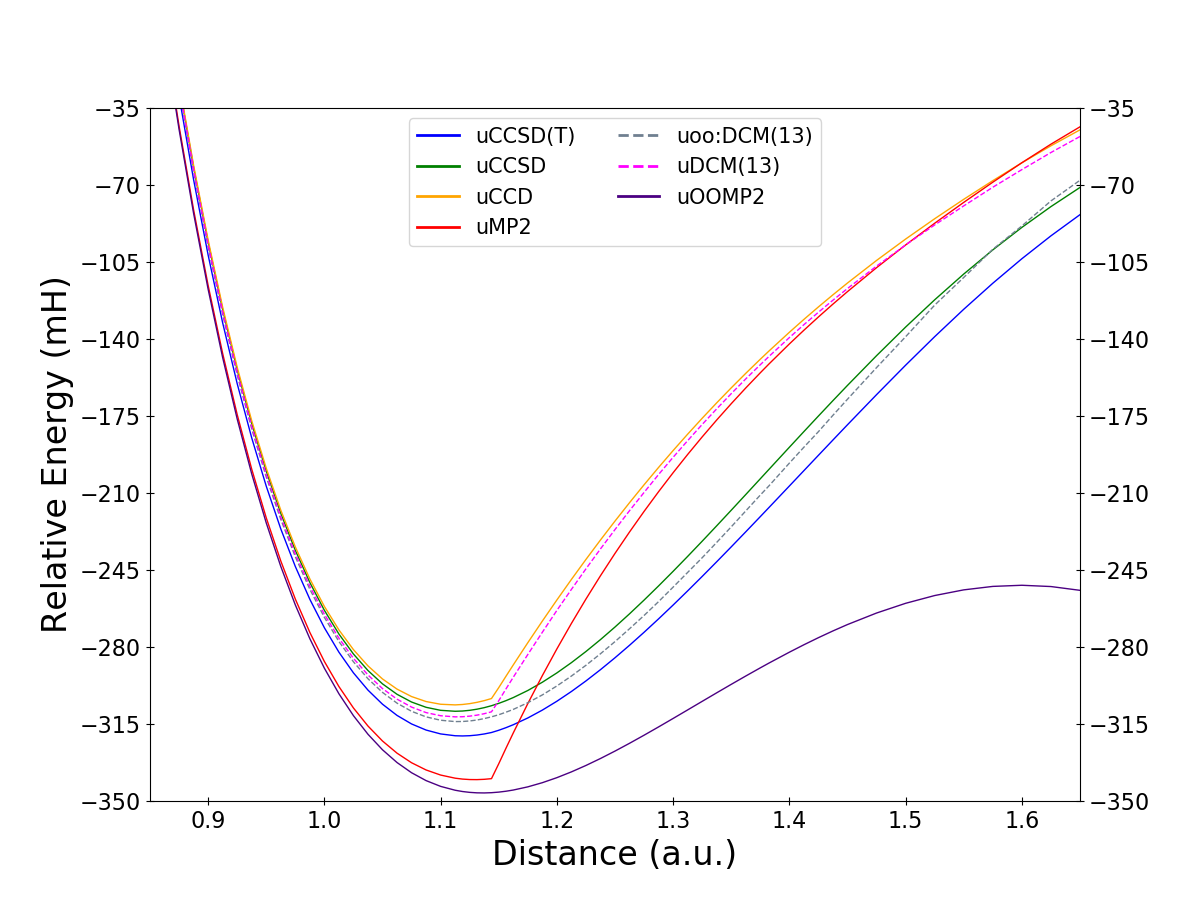}
\caption{Approximations to the total energy of N$_{2}$ across the potential curve with a UHF reference for correlation methods including uDCM(13) and uooDCM(13) as high order representatives of the uDCM(N) sequence. Energies are reported relative to two $^4$N atoms for each method.}
\label{fig:oon2}
\end{figure}

As other work has recently shown, the inclusion of an orbital-optimized reference delays the onset of the Coulson-Fischer point and one may recover better qualitative description within the well. Combined with the performance in energetics, this makes uoo:DCM(N) an excellent candidate to evaluate in the same way. 
As the uDCM(13) is the lowest order to reach the convergence limit, we now focus on only this order for both sets of orbitals to make the trend more apparent. 
Figure \ref{fig:zoom_oon2} shows results similar to above near the well, while \ref{fig:oon2} shows the behavior up through the qualitative failure and through the point where the uOOMP2 curve turns over (note that UOOMP2 does not exhibit a Coulson-Fischer point, as has been extensively discussed elsewhere\cite{Stuck:2013,Lee:2018a}). The uoo:DCM(N) curve does in fact recover the correct qualitative behavior near equilibrium and well after Hartree-Fock's spin-polarization transition. 

Interestingly, even as the OOMP2 curve begins to turn over, the uoo:DCM(13) results are able to delay this to be more in line with the uCC methods. By the time the former reaches its maxima, this is no longer the case and both begin to diverge after this point. However, it is notable that the moments approach delays the turning point as well. Of course, while the the uDCM(13) does have its discontinuity, the curve does not divergence. Additionally, uoo:DCM(13) is able to surpass quantitatively all methods here relative to CCSD(T), now resting at 6.4 mH away rather than CCSD's 11.2 mH. 

\section{Conclusions}

The purpose of this work was to revisit the connected moment expansion (CMX) in order to introduce an approximation analogous to those that have been successful in widely used single-reference correlation methods such as coupled cluster theory. The exact moments (powers of the Hamiltonian) that enter the CMX expressions have compute costs that rise incredibly steeply with the power, which makes these otherwise attractive non-iterative methods unfeasible in practice.
To circumvent the computational bottleneck, we decided to evaluate the moments using only the doubles part of the Hilbert space that is strongly orthogonal to a single reference such as the Hartree-Fock determinant, or an approximate Brueckner determinant. We call this approach the Doubles Connected Moments (DCM) expension, and, via the CMX framework, it gives rise to a tractable sequence of DCM energies, DCM(N), for $\mathrm{N}=1 \cdots \infty$. Beyond $N=1$, each DCM(N) energy can be evaluated with compute cost that scales the same as an iteration of the CCD or CCSD equations. The 
fact that DCM(N) energies are constructed from connected moments allows the method to retain the important property of size-extensivity of the exact CMX(N) energies.

The DCM(N) methods have been implemented to employ single-references that can be either Hartree-Fock or approximate Brueckner from orbital-optimized MP2 (oo), and can be either spin-restricted or unrestricted. The resulting methods were then assessed on the correlation energies of the 55 small molecules (and radicals) in the G1 data set, against virtually exact results from selected configuration interaction, as well as standard MP2, CCD, CCSD, and CCSD(T). Interestingly, we observe that DCM(N) for $\mathrm{N}>10$ performs quantitatively well relative to infinite order doubles methods such as CCD and CCSD. These are perhaps the first tractable calculations using the moments approach that appear potentially viable for chemical applications. Statistically, the oo:DCM(N) energies outperform CCSD for the correlation energy, while the Hartree-Fock-based DCM(N) energies outperform CCD for the correlation energy. Examination of both individual and collective results shows generally smooth convergence patterns to limiting values by DCM(11-14).
The DCM(N) methodology looks useful already although we stress that it is important to explore questions of numerical stability, and basis set extension carefully in future work. If those results are positive, as seems quite likely, then there are practical ways in which the methodology can be systematically improved. First, one may refine the approximation of the moments to include further contributions. In particular, one can easily imagine the construction of SDCM(N) by including the singles contributions to $\mu_{2}$ and higher. 
Additionally, there are opportunities to further optimize the implementation. For example, the most costly step in DCM(N) is the contraction of the particle ladder terms, and the use of the tensor hypercontraction (THC) formalism would bring the scaling of DCM(N) down to $\mathcal{O}(M^{4})$ with molecule size, as in the case of MP3. \cite{lee2019systematically} There are likewise opportunities to consider more efficient implementations of higher connected contributions such as triples. For example, the $\mu_{4}$ connected triples may be calculated in $O(M^{6})$ time unlike its analogue in the MP4 case where a 6-index denominator results in its most expensive contribution scaling as $O(M^{7})$. Though this specific contribution is still relatively intractable at $O(v^{6})$ (where $v$ is the number of virtual orbitals), the potential factorizability of the method provides further opportunities for use of THC and Resolution-of-the-Identity (RI). 

Especially in situations that exhibit an interplay of strong and weak correlation, the DCM(N) approach may be a blueprint for further refinement of more sophisticated yet still low-cost references, such as strongly orthogonal geminal wavefunctions\cite{hurley1953molecular} or the coupled cluster valence bond (CCVB) reference.\cite{Small:2009,small2011post} Indeed the prior success seen using low-order CMX with APSG and CAS reference wavefunctions suggests there is potential for new progress in this direction, based on retaining the affordability and promising accuracy that is a key feature of DCM(N).


\section{Acknowledgments} 
This work was supported by the Director, Office of Science, Office of Basic Energy Sciences, of the U.S. Department of Energy through the Gas Phase Chemical Physics Program, under Contract No. DE-AC02-05CH11231.  BG acknowledges additional support from the National Science Foundation Graduate Research Fellowship under Grant No. DGE 110640 and DGE 1752814 and the UC Berkeley Chancellor's Fellowship.

\bibliography{moment}

\end{document}